\documentclass[11pt,cmr,a4paper,epsf]{article}
\usepackage{moriond,epsfig}
\def\lyal{$Ly\alpha$}

\def\be{\begin{equation}}
\def\ee{\end{equation}}
\def\bea{\begin{eqnarray}}
\def\eea{\end{eqnarray}}

\def\apj{ApJ}%
\def\apjl{ApJ}%
\def\mnras{MNRAS}%
\begin{document}
\vspace*{4cm}

\title{The Fate of Lyman photons in local Starburst: new ACS/HST images}
\author{ {\bf Daniel Kunth} (1), M. Hayes (2), G. \"Ostlin (2),
M.J. Mas--Hesse (3), C. Leitherer (4), A. Petrosian (5)}
\address{(1) Institut d'Astrophysique de Paris
98 bis Bld Arago, 75014 Paris, France}
%
% \author{M. Hayes}
%
\address{(2) Stockholm Observatory, SE-106 91 Stockholm, Sweden}
%
% \author{G. \"Ostlin}
%
% \address{Stockholm Observatory, SE-106 91 Stockholm, Sweden}
%
% \author{M.J. Mas--Hesse}
%
\address{(3) Centro de Astrobiologia (CSIC--INTA), Madrid,  Spain}
%
% \author{C. Leitherer}
%
\address{(4) Space Telescope Science Institute, 3700 San Martin Dr., Baltimore, MD
21218,USA}
%
% \author{A. R. Petrosian}
%
\address{(5) Byurakan Astrophysical Observatory, Byurakan 378433,
Armenia}
\maketitle

\abstract{We review the imaging and spectroscopic properties of the \lyal\ 
emission  in
starburst galaxies. The \lyal\ photons escape is largely driven by 
kinematical and orientation effects related to the presence of  
large scale expanding shells of neutral hydrogen. The various  
\lyal\ profiles can be linked to the evolutionary state of the starburst. 
Our recent \lyal\ imaging study using the 
HST-ACS is presented here: it highlights the presence of diffuse extended
emission of low intensity leaking out the diffuse HI components. The
similarities with the line profiles observed on high redshift galaxies is an
important fact that cautions the use of $Ly\alpha$\  to derive the cosmic 
star formation rate and constrains the search for 
high redshift galaxies.}

\section{Introduction}

        Star-formation rates in primeval galaxies are expected to reach 
hundreds of $M_\odot$yr$^{-1}$ (Partridge \& Peebles,1967).  For a normal Salpeter IMF 
this corresponds to total
bolometric luminosities in excess of $10^{11}L_\odot$, which is similar to 
the values in luminous IRAS galaxies (Heckman 1993). The ionizing radiation 
from the 
newly formed young stars should lead to prominent \lyal\ emission due to
 recombination of hydrogen in the ambient interstellar medium. Therefore,
 the \lyal\ line could be an important spectral signature in young galaxies
 at high redshift since the expected \lyal\ luminosity amounts to a few
 percent of the total galaxy luminosity (Schaerer 2003; Stiavelli et al. 2003).
        
        From the above estimates, typical \lyal\ fluxes of 
$10^{-15}$~erg~s$^{-1}$~cm$^{-2}$\ are expected. Such values are within
 easy reach of present instruments. Over the past 20 years major 
observational efforts were undertaken to search for \lyal\  emission from
 faint galaxies at high redshift (Djorgovski \& Thompson 1992). Although 
many \lyal\ emitters have been found (e.g., Frye et al. 2002;
 Malhotra \& Rhoads 2002; Fujita et al. 2003; Ouchi et al. 2003, Schaerer,
this conference),
 their numbers are smaller than predicted. Where is the population of
 \lyal\ emitting field galaxies, which should exist at high redshift?

The assumption of the \lyal\ intensity as produced by pure recombination in
a gaseous medium may be too simple. Meier \& Terlevich 1981, 
Hartmann et al. 1988, 
Neufeld 1990, and Charlot \& Fall 1993 considered the
effects of dust on \lyal. \lyal\ photons experience a
large number of resonant scatterings in neutral atomic hydrogen, thereby
increasing the path length and the likelihood of dust scattering and
absorption. This process can be very efficient in removing \lyal\ photons
from the line of sight to the observer, leading to much lower line
strengths in comparison with the idealized Case B. Depending on the aspect
angle of the galaxy as seen from the observer, this may lead to a decrease
of the \lyal\ equivalent width. On the other hand, \lyal\ may actually be
enhanced due to the presence of many supernova remnants which form during
the starburst (Shull \& Silk 1979). The net result is
controversial. Bithell (1991) finds supernova remnants to be an important
contributor to the \lyal\ strength whereas Charlot \& Fall (1993) reach the
opposite conclusion.

The theoretical situation is sufficiently complex that observational tests
are required. The most obvious test are measurements of \lyal\ in local
starburst galaxies whose redshifts are sufficiently large to permit
observations of their intrinsic \lyal\ outside the geocoronal and Galactic
interstellar \lyal. Observations of local starbursts have indeed been
performed with the {\it IUE} satellite, (Meier \& Terlevich 1981;
Hartmann et al. 1988; Calzetti 2001; 
Terlevich et al. 1993; Valls-Gabaud 1993).
Again, the results are controversial. For instance, Calzetti \& Kinney (1992) 
and
Valls-Gabaud find \lyal\ strengths in agreement with pure recombination
theory whereas Hartmann et al. and Terlevich et al. conclude that
significant dust trapping of \lyal\ photons must occur.

\section{The role of kinematics in HI gas}
 
Finally, and most importantly, the kinematic properties of the 
interstellar medium may very well be the dominant escape or trapping 
mechanism for \lyal\  (Kunth et al. 2003).
The complex nature of the \lyal\ escape probability has been revealed by 
the HST/GHRS spectroscopy (Kunth et al. 1998) and raised additional 
questions.  They detected $Ly\alpha$\ emission in half of them,
with clear asymmetric P--Cyg profiles, as shown in  Fig.~1a, while
the other half showed prominent damped $Ly\alpha$\ absorptions. The sample was
extended at lower resolution by Thuan \& Izotov (1997). In Fig.~1b
we show the 3 characteristic $Ly\alpha$\ profiles that can be found in
star-forming galaxies: pure emission with symmetric profile, P--Cyg emission
and broad, damped absorption profile.

The analysis of these data yielded the following results:

\begin{figure}
\includegraphics[width=5cm]{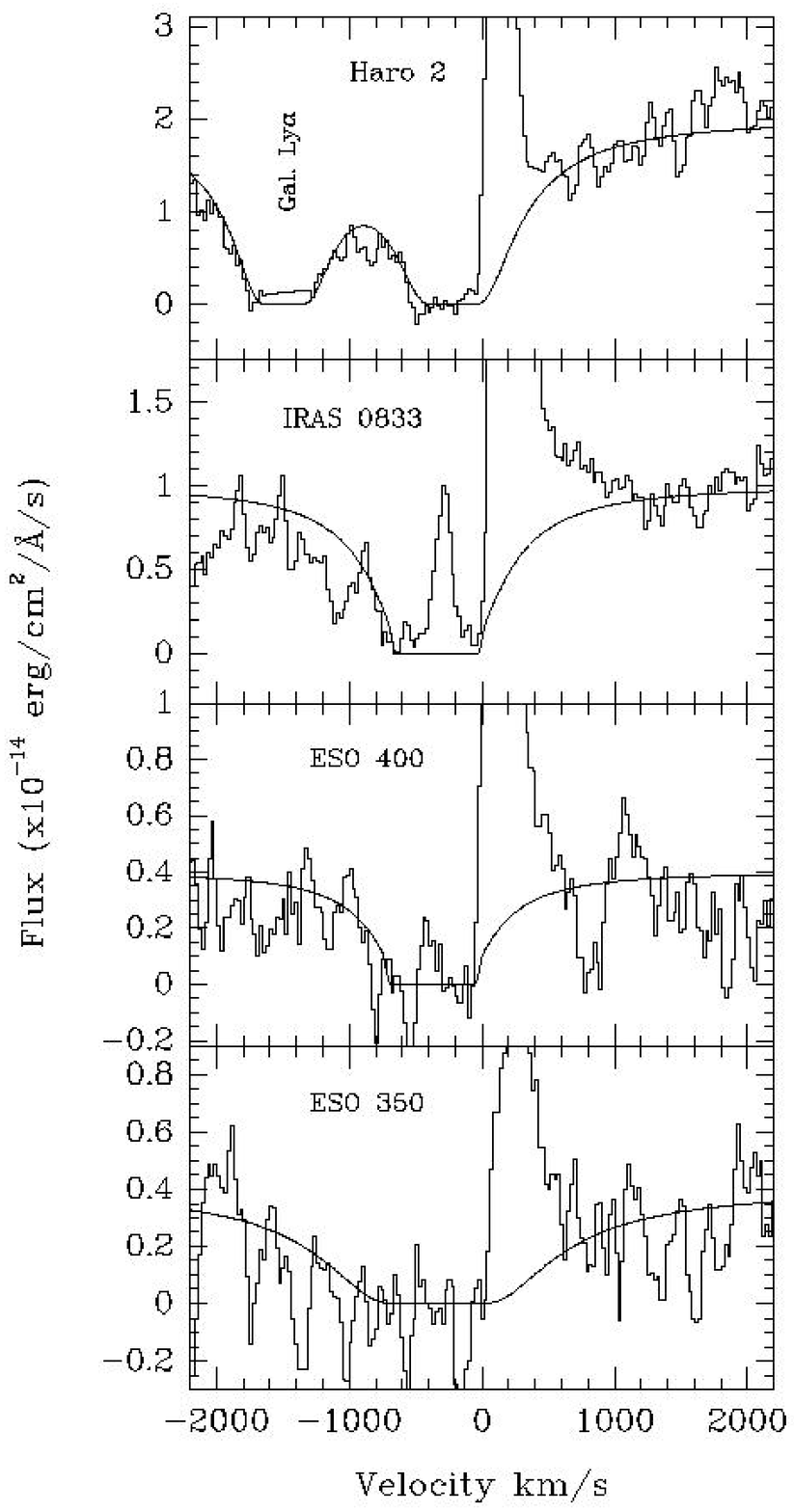}
\includegraphics[width=5cm]{mmf1b.eps}
\centering
\caption{a) Starburst galaxies showing $Ly\alpha$\ emission lines with P--Cyg
  profiles, as shown by Kunth et al. (1998); b) prototypical examples of
  the three cases discussed in the text: pure emission, P--Cyg profile and
  damped absorption.} 
%\label{f1}
\end{figure}

\begin{figure}
\includegraphics[width=5cm]{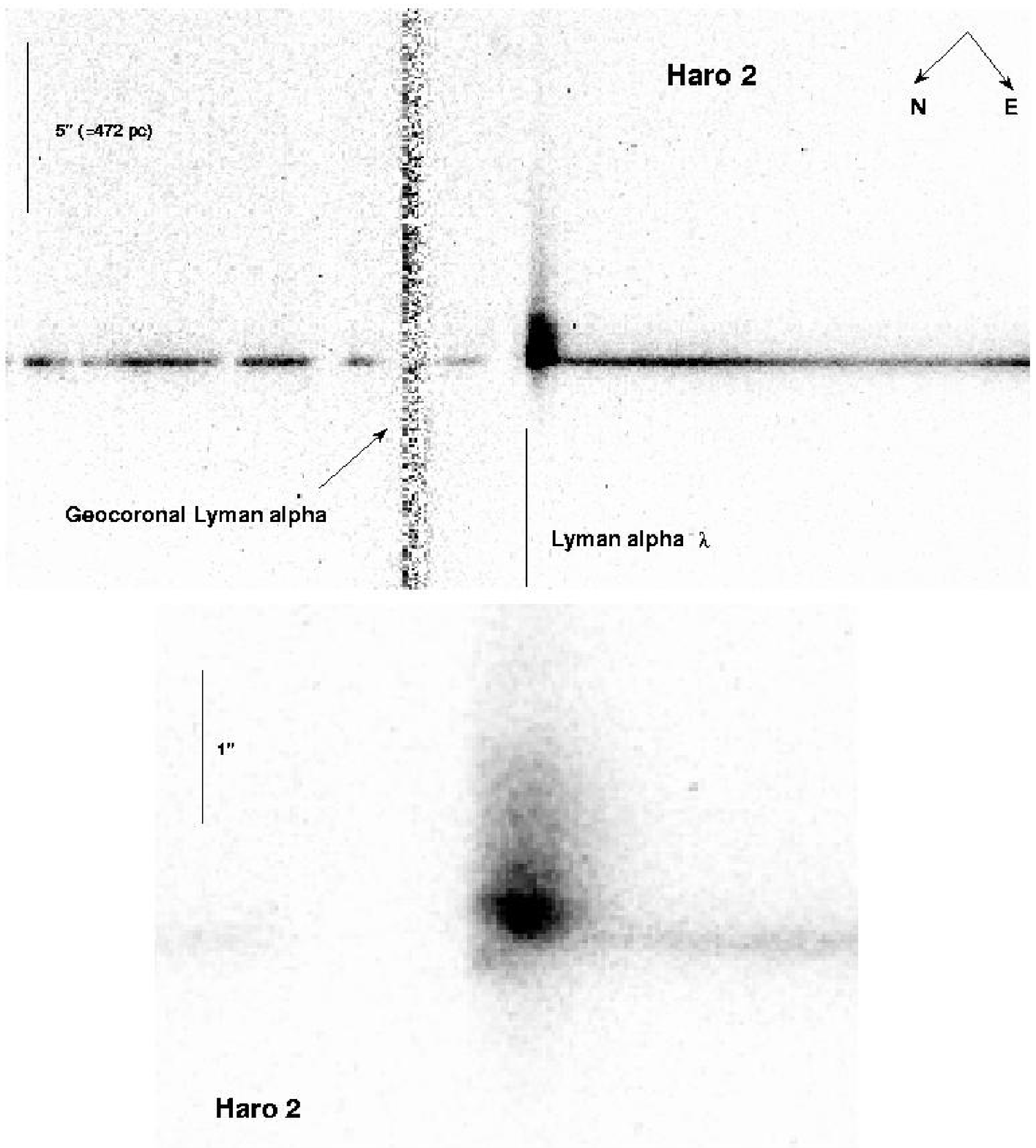}
\includegraphics[width=5cm]{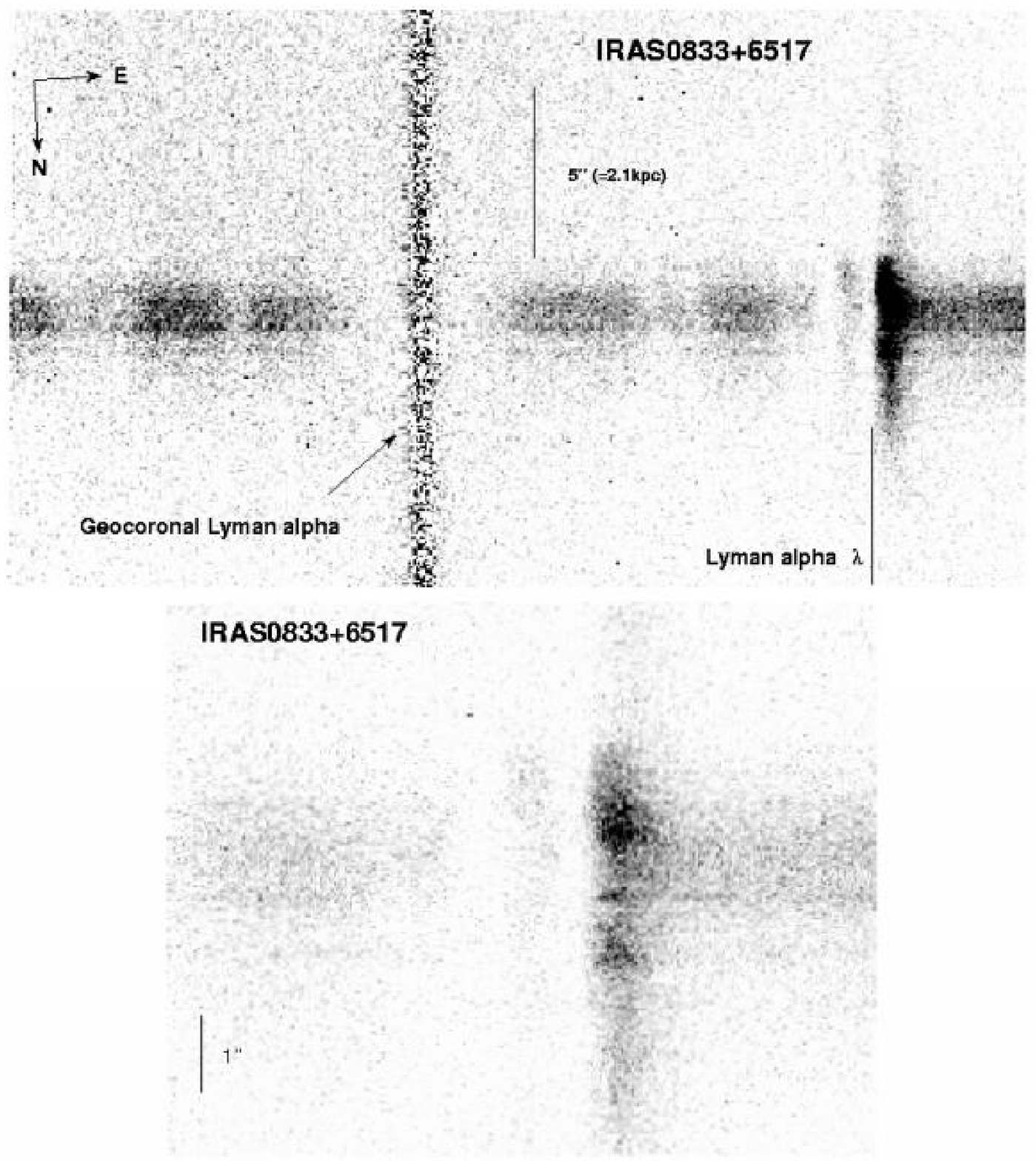}
\centering
\caption{a) Spectral HST--STIS image of Haro~2 around the $Ly\alpha$\ region. 
  Wavelength
  increases upwards. The $Ly\alpha$\ emission is extended over around 8 arcsec\ 
  (1 kpc). Note the extended low density component and the sharpness of the
  blue absorption edge. b) Spectral image of IRAS~0833+6517.  The $Ly\alpha$\
  emission 
  is extended over around 10 arcsec (5 kpc). Note the presence of a
  secondary emission peak located
  at the center of the absorption profile. }
%\label{f2}
\end{figure}

\begin{itemize}

\item As $Ly\alpha$\ is detected in emission,  a
clear P--Cyg profile is seen in most cases. Also the interstellar
 neutral metallic lines appear blueshifted by 200--400 km/s.

\item When the neutral gas is static with respect to the HII region, a
  damped broad absorption profile is detected.  

\end{itemize}

Hence HST-GHRS data clearly  showed  that the kinematics of the gas was one
major parameter determining the visibility of the $Ly\alpha$\ line. 
 However, 
if the dust content is small, the \lyal\ photons may, after multiple resonant 
scatterings could diffuse out over a larger area. This would create a bias in 
the spectroscopic studies which usually target the regions of peak UV 
intensity - under this scenario the places where we do {\em not} expect 
to see \lyal\ in emission. Another
possibility is that the UV-continuum sources are partly shielded by a clumpy
medium, in which case we would see mixed absorption and emission. In cases
where the ISM has a non-zero radial velocity with respect to the UV continuum
source, \lyal\ may appear in emission with a characteristic P-Cygni profile.
Depending on the morphology and kinematics of the galaxy this can occur in
different regions, and e.g. in a galaxy merger the escape probability may be 
enhanced.

Further on, Mas-Hesse et al. (2003) have
performed 2--D observations using HST-STIS in order to analyze the
spatial structure of the emission profile, and look for areas where the
$Ly\alpha$\ photons could be leaking. Three objects were
included in the sample: 2 showing a strong P--Cyg profile (Haro~2 and
IRAS~0833+6517) and 1 showing a damped absorption (IZw18). 

No emission whatsoever was detected from IZw18 along the STIS slit. 
The spectral
images for Haro~2 and IRAS~0833+6517 are shown in Fig.~2. The main results
drawn from these data are: 

\begin{itemize}

\item No velocity structure has been detected on the sharp edge of the 
P--Cyg profiles on scales of 1--5 kpc. 

\item These profiles imply the presence of large column densities of
  neutral gas outflowing from the HII region at velocities of 200 -- 400
  km/s, acting practically as a moving plane-parallel slab on kpc scales. 

\item Detection of broad and extended emission
  components of low intensity. 

\end{itemize}  

%\section{Interpration: an evolutive view}
\section{Interpretation: an evolutionary view}

These observational results suggest that the central starbursts in these
galaxies are driving a huge expanding shell of neutral gas. One can postulate
that the interaction of the ionizing flux and the shell itself with the
surrounding medium can lead to different configurations which could explain
the variety of $Ly\alpha$\ profiles. Moreover, these scenarios could be
correlated with the evolutionary state of the starburst process. 

Mas-Hesse et al. (2003)  show first in Fig.~3 the predicted effect of an 
expanding shell of neutral Hydrogen for different column densities. 
P--Cyg profiles similar to the observed ones are produced for column 
densities $\log(n_H) \approx 19 - 20 $ cm$^{-2}$, moving at velocities 
around 300 km/s with respect to the
HII region where the $Ly\alpha$\ emission line is produced. For higher
column densities the absorption is completely damped, and no $Ly\alpha$\
photons can escape. On the other hand, for much higher outflowing
velocities, the absorption takes place so much to the blue of the emission
profile, that this wouldn't be affected at all. Note that
as a result of this saturated absorption by neutral gas, the centroid of
the emerging $Ly\alpha$\ emission line might appear artificially redshifted
by a significant amount (see Fig.~3b) with respect to the HII region. This
effect can be mistaken as an evidence for the presence of receding flows of
ionized gas, which is not the case  since the line is indeed 
emitted by the HII region.

\begin{figure}
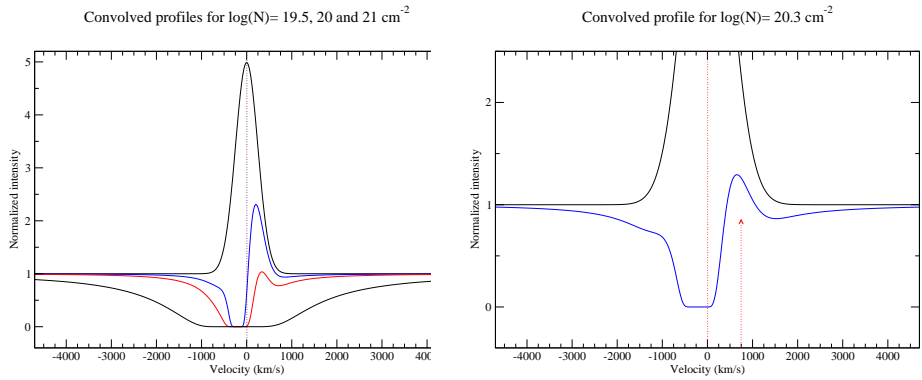

\includegraphics[width=6cm]{mmf3a.eps}
\includegraphics[width=6cm]{mmf3b.eps}
\centering
\caption{Effects of an expanding shell on the $Ly\alpha$\ emission
  profile. The left panel shows the intrinsic emission line, and the
  resulting profiles for different column densities of neutral Hydrogen. On
  the right panel we show that the resulting emission might appear
  artificially redshifted by several hundreds of km/s. }
%\label{f3}
\end{figure}

As discussed in detail in Tenorio-Tagle et al. (1999)  and Mas-Hesse et
al. (2003), the interaction of bubbles and superbubbles as the starburst
episode evolves may lead to different scenarios affecting the properties of
the $Ly\alpha$\ emission line profiles. We have identified 4 basic steps: 

\begin{itemize} 

\item[i)] Initially, when a star-forming episode starts, a central HII
region begins to develop. At this phase, if the neutral gas surrounding the
starburst region has HI column densities above 10$^{14-15}$ cm$^{-2}$, an
absorption line centered at the systemic velocity of the galaxy will be
visible, independently on the viewing angle. If the total HI column density
along the line of sight is higher than around 10$^{18}$ cm$^{-2}$, 
a damped $Ly\alpha$\ 
absorption profile will be detectable. It is important to stress that
during this early phase of a starburst the Balmer emission
 lines will be strongest,
due to the high ionizing flux produced by the most massive stars.

\item[ii)] The situation changes drastically and becomes a strong function
of viewing angle, once the mechanical energy released by the starburst is
able to drive a shell of swept up matter exceeding the dimensions of the
central disk. Then, upon the acceleration that this shell experiences as it
enters the low density halo, it becomes Rayleigh--Taylor unstable and
fragments.  This event allows the hot gas (composed basically by matter
recently processed by the starburst), to stream with its sound speed
between fragments and follow the shock which now begins to form a new shell
of swept--up halo matter, expanding at a velocity $v_{exp}$.
 Another consequence of the blowout is the fact
that the ionizing photons from the recent starburst are now able to
penetrate into the low density halo, and manage to produce an extended
conical HII region that reaches the outskirts of the galaxy. 
An observer looking then at the starburst through the conical HII region
will be able to detect the strong $Ly\alpha$\ emission line produced by the
central HII region, centered at the systemic velocity of the galaxy, and
without any trace of absorption by neutral gas.  On the other hand, an
observer looking far away from the conical HII region will detect a broad
absorption profile at any evolutionary state.

\item[iii)] Sooner or later, recombination will begin
in the expanding shell. This will cause a strong depletion of the ionizing
radiation which formerly was able to escape the galaxy after crossing the
extended conical HII region. Recombination in the expanding shell will
produce an additional broad $Ly\alpha$\  component of low intensity.  

\item[iv)] The ionization front becomes eventually trapped within the
expanding shell by basically 3 effects. First, by the increasingly larger
amount of matter swept into the expanding shell, as this ploughs into the
halo. Second, the growth of the shell dimensions also implies less UV
photons impinging, per unit area, at the inner edge of the shell. And
third, in the case of a nearly instantaneous starburst, the production of
UV photons starts to decrease drastically (as $t^{-5}$) after the first 3.5
Myr of evolution.

The trapping of the ionization front will lead to the formation of a neutral
layer at the external side of the expanding shell.  All these effects result
in an increasingly larger  saturated absorption, as the external
neutral layer will resonantly scatter the $Ly\alpha$\  photons.
This absorption will appear blueshifted with respect to the $Ly\alpha$\ 
emitted by the central HII region by $-$$v_{exp}$\, leading so to the formation
of a P--Cyg profile where a variable fraction of the intrinsic $Ly\alpha$\ 
emission would be absorbed.   

In addition, the profile will be contributed by the $Ly\alpha$\  
radiation arising
from the receeding section of the shell, both by recombination on the
ionized layer, and by backscattering of the central $Ly\alpha$\  photons by the
neutral layer. 

\end{itemize}

Under some circumstances, the leading shock front on the external surface
of the shell can be heated and become ionized, producing so an additional
$Ly\alpha$\ emission which would be detected blueshifted by $-$$v_{exp}$\, 
and not
affected by absorption. This could be the case of the secondary emission
peak detected in IRAS~0833+6517. 

\section{The ACS imaging studies}

In the distant universe, \lyal\ imaging and low resolution spectroscopic
techniques are now successfully used to find large numbers of galaxies.
However, without a proper understanding of the \lyal\ emission processes
this line cannot be used to estimate astrophysical quantities such as star
formation rates and fluxes for reionisation and it becomes
dubious to use it to study clustering if the biases are not properly known. 
If the star forming activity of a high redshift galaxy is connected with 
its environment, the \lyal\ escape probability will not be independent of
this parameter.

\medskip

These considerations led us to start a pilot programme to image {\em local} 
starburst galaxies in the \lyal\ line using the solar blind channel (SBC)
of the Advanced Camera for Surveys (ACS) onboard HST, allowing us to study
the \lyal\ emission and absorption morphology in detail. A sample of six
galaxies with a range of luminosities ($M_V = -15$ to $-21$) and 
metallicities ($0.04 Z_\odot {\rm ~to} \sim Z_\odot$), including previously
known \lyal\ emitters as well as absorbers, were selected and observed during 
30 orbits in Cycle 11. The observations were obtained through the F122M 
(\lyal\ ) and F140LP (continuum) filters.

\begin{table}[h]
\caption[]{Targets for ACS \lyal\ imaging project 
%{\tt If we want to have this table, we should check the numbers below}
} 
%\begin{flushleft}
\begin{center}
\begin{tabular}{lccc} 
\hline
Galaxy	& $M_B$ &  12+log(O/H) & emitter/absorber \\
\hline 
\noalign{\smallskip}
SBS\,0335-052	& --17 & 7.3 & absorber \\	
NGC\,6090	& --21 & 8.8 & emitter  \\
ESO\,350-38	& --20 & 7.9 & emitter  \\
Tol\,1924-416	& --19 & 7.9 & emitter  \\
Tol\,65		& --15 & 7.6 & absorber \\
IRAS\,08+65	& --21 & 8.7 & emitter \\ 
\hline
\end{tabular}
%\end{flushleft}
\end{center}
\end{table}

\subsection{First continuum subtractions}

The first results for two of the galaxies, ESO\,350-38 and 
SBS\,0335-052, were presented in Kunth et al. (2003). The images were 
drizzled to correct for geometric distortion, aligned and background
subtracted. In order to subtract the continuum (F140LP) from the
on-line (F122M)
images to construct line-only \lyal\ images it is necessary to assume 
a shape of the continuum. As the UV spectra of starbursts are fairly
well described by power-law spectral energy distributions (SEDs), a 
natural first step would be to adopt a power-law, $f_{\lambda}=
\lambda^{\beta}$, with the slope $\beta$ derived from, e.g. IUE spectra 
in the range $\lambda \sim 1300$\AA . The relative scaling factors between
the filters differs by a factor 1.7 for assumptions of $\beta =-2$ and $1$. 
The \lyal\ images in Kunth et al. (2003) were obtained in this way, 
by assuming $\beta =0$, however it was noted that $\beta$ was variable 
over the face of each galaxy. 

\medskip
  
The results presented in Kunth et al. (2003) give a view of \lyal\  that 
is complementary to the results of previous HST/GHRS spectroscopy by
Kunth et al. (1998). Moreover, they reveal the complex nature of
\lyal\  emission and absorption in starburst galaxies. ESO350-IG038 
shows \lyal\  in emission from several knots (A and C following
Vader el. al 1993). By using the F140LP image,
and an archival F606W WFPC2 image these knots were found to 
have very blue UV/optical colours 
($f_{\lambda , F140LP} / f_{\lambda , F606W} = 15 $). In knot B and the 
surrounding region, \lyal\  is seen in absorption. Here, colours are much 
redder ($f_{\lambda , F140LP} / f_{\lambda , F606W} = 1 $). Knot A also
shows a P Cygni profile and and diffuse \lyal\  emission is seen in 
numerous regions, particularly to the south-west of the image. 

%	{\tt NB Names of different knots and SSCs in ESO350 and SBS
%	should only be discussed if we actually show a figure of them.}

\medskip

The  image of SBS0335-052 shows broad damped \lyal\  absorption almost
throughout, confirming the reports of Thuan et al. (1997).
Diffuse \lyal\ 
emission is seen towards the north of the image (around SSC 5 following
Thuan et al.) although it is very sensitive to 
the assumptions in the continuum subtraction. 

\medskip

Clearly, in order to separate emission from absorption in the less obvious
cases, and in order to obtain photometrically valid images, a continuum
subtraction procedure that takes the spatial variation of the continuum 
slope into account, is necessary. 

\begin{figure}[h]
\resizebox{0.5\hsize}{!}{\includegraphics{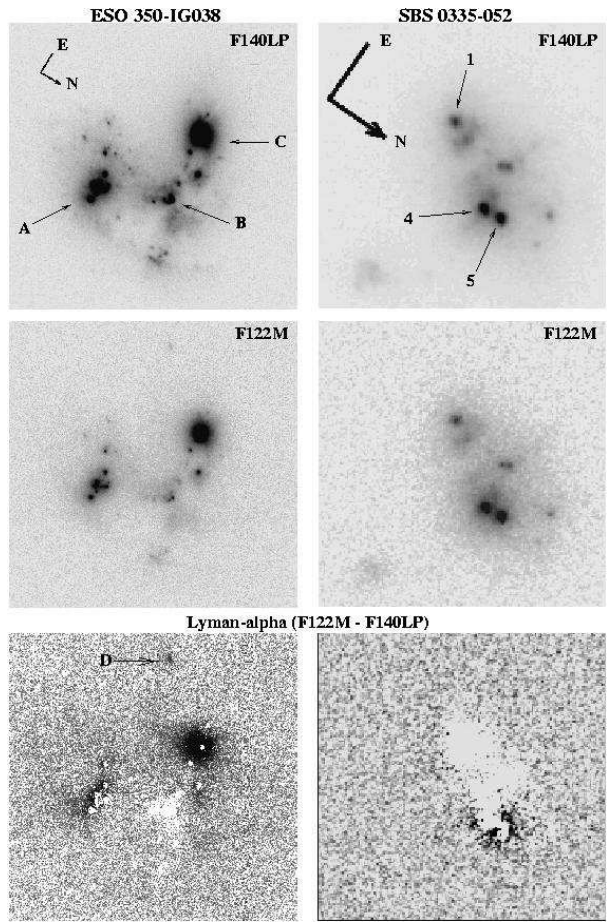}}
%   \sidecaption
\centering
\caption{{\bf Top:} UV continuum (F140LP) images of ESO\,350-38 (field shown
is 13 arcsec x  13 arcsec) and SBS\,0335-052 (4.4 arcsec x 
4.4 arcsec). The different knots discussed in the text are labeled.  
{\bf Middle:} F122M images.  {\bf Lower panels:} continuum subtracted \lyal\ 
 image
assuming $\beta =0$. Emission is shown in black and absorption in white.
Note the faint \lyal\  emitting blob (labeled D) in ESO\,350-38 and the
possible faint emission to the north of '4' and '5' in SBS\,0335-052. }
  \label{FigFirstTwo}
\end{figure}

\subsection{Continuum subtractions using a $\beta$-map}

A more realistic way of estimating the scaling factor would be to use 
photometric maps of the ultraviolet continuum slope, so called
``$\beta$-maps''. 
A $\beta$-map can be constructed from the F140LP images and an HST image
in another, preferably blue, passband. Many of the target galaxies have
been previously imaged e.g. with WFPC2 and $\beta$-maps were constructed
based on these. A general feature of such continuum subtractions is that
the continuum is oversubtracted. This calls for several explanationss: 
i) The continuum often starts to deviate from a power-law at wavelengths 
$\lambda < 1300$ \AA, and in addition, absorption features may be 
present.
Moreover, this $\lambda$-region is sensitive to extinction and the
the extinction near \lyal\ is greater than for the continuum.
ii) The F122M filter also covers, in addition to \lyal\ in the target 
galaxies, \lyal\ absorption from gas in the Milky Way. This means 
that a fixed fraction of the photons with wavelength $\lambda_{\rm obs}
= 1216 \AA  \pm \delta \lambda$ will be lost. Since the solid angle covered
is very small, this fraction will be fixed for each galaxy and will
depend only on the column density of neutral hydrogen along the
sightline.
        
\medskip

It should be pointed out that the first point concerns also
high-$z$ \lyal\ galaxies where continuum subtraction is a
potential, yet little discussed, problem. In addition there
might be an effect of intergalactic dust along the sightline.
%{\tt (check this)}. 
What is needed is
a way to relate the correct continuum level at \lyal\ to the
colours at longer wavelength.

\medskip

We proceeded by switching our attention to another 
galaxy in our sample, ESO338-IG04 (Tololo 1924-416) where we have
deep HST/WFCP2 images in the F218W, F336W, F439W, F555W and F814W
filters (\"Ostlin et al. 1998, 2003),  as well as STIS
long slit spectra with G140L available (Leitherer et al. in preparation). 
%{\tt (check this with Claus)}
 The combination of spectroscopic 
and imaging data over a wide wavelength interval meant that this
target was ideal for a thorough investigation of the continuum 
subtraction procedures. From IUE spectra (e.g. Giavalisco et al.
1996) this target was known to be one of the
brighter \lyal\ emitters in the local universe.

\medskip

\begin{figure}[h]
\resizebox{0.7\hsize}{!}{\includegraphics{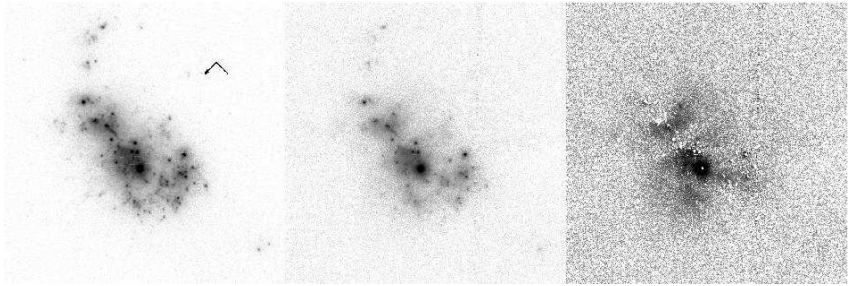}}
%   \sidecaption
\centering
\caption{{\bf Left:} UV continuum (F140LP) image of ESO\,338-IG04.
             The different knots discussed in the text are labeled.  
         {\bf Middle:} F122M image.  
         {\bf Right:} continuum subtracted \lyal\  image assuming $\beta =0$.
                Emission is shown in black and absorption in white.
               The field shown is 14 arcsec x 14 arcsec. }
         \label{FigE338_3}
\end{figure}

Preliminary continuum subtractionsusing a flat 
($\beta = 0$) continuum, similar to those
produced for ESO350-IG038 and SBS0335-052 had already shown
ESO338-IG04 to be a bright \lyal\  emitter over nearly all of 
the starburst region. This can be seen in Figure ~\ref{FigE338_3}.
Small offset absorption holes appeared in the brightest regions, similar
to those announced for ESO350-IG038 in Kunth et al. (2003) and a 
feature consistent with a P-Cygni profile was visible in the 
brightest central star cluster. 
Also, a lot of diffuse \lyal\  was seen leaking out from regions 
approximately east and west of the central starburst. 
Adopting the normal $\beta$-map procedure again lead to the 
usual oversubtraction of the continuum, even after experimenting
with $\beta$-maps constructed from a variety of filter combinations. 

\medskip

Comparison with the low resolution STIS
spectrum  of the central regions proved this continuum subtraction to 
be inaccurate. The spectrum shows \lyal\  emission from regions were 
the $\beta$-map subtracted image shows absorption.  Comparison of the
$\beta$ values from the images and spectra in several star forming 
regions revealed that photometrically determined $\beta$ values were 
usually more 
negative than those determined by fitting a power law to the slope of 
the spectrum. 
It is clear that 
in at least some cases, the method of continuum subtraction using 
$\beta$-maps substantially underestimates the scaling factor between 
the F122M and F140LP filters, leading to the subtraction of too much 
continuum. 

\medskip

The method of using a $\beta$-map
to estimate the scaling factor indeed relies upon the power-law being 
continuous at wavelengths $\lambda < 1400$\AA. The IUE spectrum of 
this galaxy seems to indicate that this is not the case.
The spectrum   seems to flatten off slightly at around 1500\AA, increasing (making 
less negative) the value of $\beta$ between out online and offline 
filters.
Moreover there is no one-to-one relationship between
the $\beta$s obtained through $\beta$-maps and power-law fitting to the 
STIS spectrum at shorter wavelengths, i.e. there is no simple relationship 
that describes the flattening of the spectrum shortwards of the 
F140LP filter.

\medskip

\subsection{Synthetic spectra techniques}

The way forward is to use synthetic UV-spectra of starbursts (these are either
based on model atmospheres or use empirical data, the latter being more
realistic
but there is a general lack of data for hot stars with low metallicities and 
there may be contaminations from interstellar absorption lines).

\medskip

%{\tt [Shorten this paragraph somewhat]}\\
High resolution synthetic stellar evolutionary spectra (M. Cervi\~no,
private communication)
were used to approximate the SED of the central cluster of the galaxy. 
Our aim was to try and recreate the observed spectra of ESO338-IG04. 
The chosen synthetic spectra  were generated using a Salpeter IMF, solar
metalicity and burst ages ranging from 3 to 6 Myr. We took the 
low-resolution STIS spectrum of the central star cluster and
fitted a Voigt profile to the wings of the damped Galactic \lyal\ absorption 
profile (giving a Galactic HI column density of 20.7 cm$^{-1}$, 
independently confirming the value of 20.8 cm$^{-1}$ from Galactic HI 
maps). This Voigt profile was then convolved with the spectrum to
simulate Galactic \lyal\  absorption. We then applied Galactic
reddening using the Cardelli law with values A$_V$=0.288
and E(B-V) = 0.087 taken from NED. 
We fitted the synthetic spectrum to the real spectrum, fitting parameters
were burst age and internal reddening using the SMC law. 
The best fit synthetic spectrum was then 
convolved with the HST/ACS/SBC F122M and F140LP instrument throughput
profiles obtained using the {\tt SYNPHOT} package in 
{\tt IRAF/STSDAS} and synthetic 
fluxes in these filters were computed. The F122M-F140LP scale factor 
required to produce a net flux of zero in a continuum subtraction was 
calculated as $scale factor = f_{F140LP} / f_{F122M}$. 
This method therefore gives us the scale factor depending only on 
the flux through the online and offline filters; totally independent of 
the shape of the slope at longer wavelengths. It is not dependent on 
the parameterisation of the continuum slope. 
For the central region, this gave a scale factor of 8.89 which 
corresponds to
a $\beta$ of -0.86. While significantly less negative than the value of 
$\beta$ = -2.08 (Calzetti et al 1994), it does represent the continuum
slope in exactly the region of interest.

\medskip

This method  works well for the bright clusters along the
STIS slit where the  starburst is young  but is less applicable as the 
burst has aged. The general problem is that the aging of a burst and 
interstellar 
extinction have very similar but non-identical effects
(reddening of the SED). 
Moreover, we need to find a method that does not rely on 
spectra, but allow us to use images. To do so we calculated a model
for the appropriate metallicity ($Z=0.001$), and Salpeter 0.1--120
$M_\odot$ IMF using the Starburst99 code (Leitherer et al. 1999).
We also included nebular emission lines through the interface with
Starburst99 to the ``Mappings'' code.
The model spectra were convolved with the WFPC2 and 
ACS/SBC filter profiles for all burst ages from 1 to 900 Myr and for
a range of reddenings from E(B--V)=0 to 0.25 using the SMC law.
We then calculated the scale factor in the above manner (taking the
Galactic \lyal\ absorption into account) for all grid points.

\medskip

We then investigated the relation between the scale factor and 
observed colours using various filter combinations, i.e. we are not primarily
interested in resolving the age--reddening degeneracy, but to get a correct
scale factor as a function of colour, whatever its reason. We quickly concluded
that a single colour index can not uniquely determine the scale factor. 
Based on our experience with $\beta$-maps and fitting of the spectra, this
was not, of course, a surprise. 

\medskip

However, when using some filter combinations involving at least two
different colour index, the scale factor can be tightly constrained.
The best combinations in this case were F140LP-F218W vs. F336W-F439W 
and F140LP-F336W vs. F336W-F439W, the latter is presented here.

The model data points were  used to populate a lookup table where the 
scale factor could be looked up from any pair of F140LP-F336W and 
F336W-F439W colours. In between evolutionary steps in the models
(jumps in burst age), more colours and scale factors were
interpolated linearly.
Using this lookup table, the ACS F140LP
image and WFPC2 F336W and F439W images were used to create a map of 
the scale factor over the whole starburst region. 
This scale factor map was then used to make a continuum 
subtraction. 

\begin{figure}[h]
\centering
 \resizebox{0.35\hsize}{!}{\rotatebox{270}
                             {\includegraphics{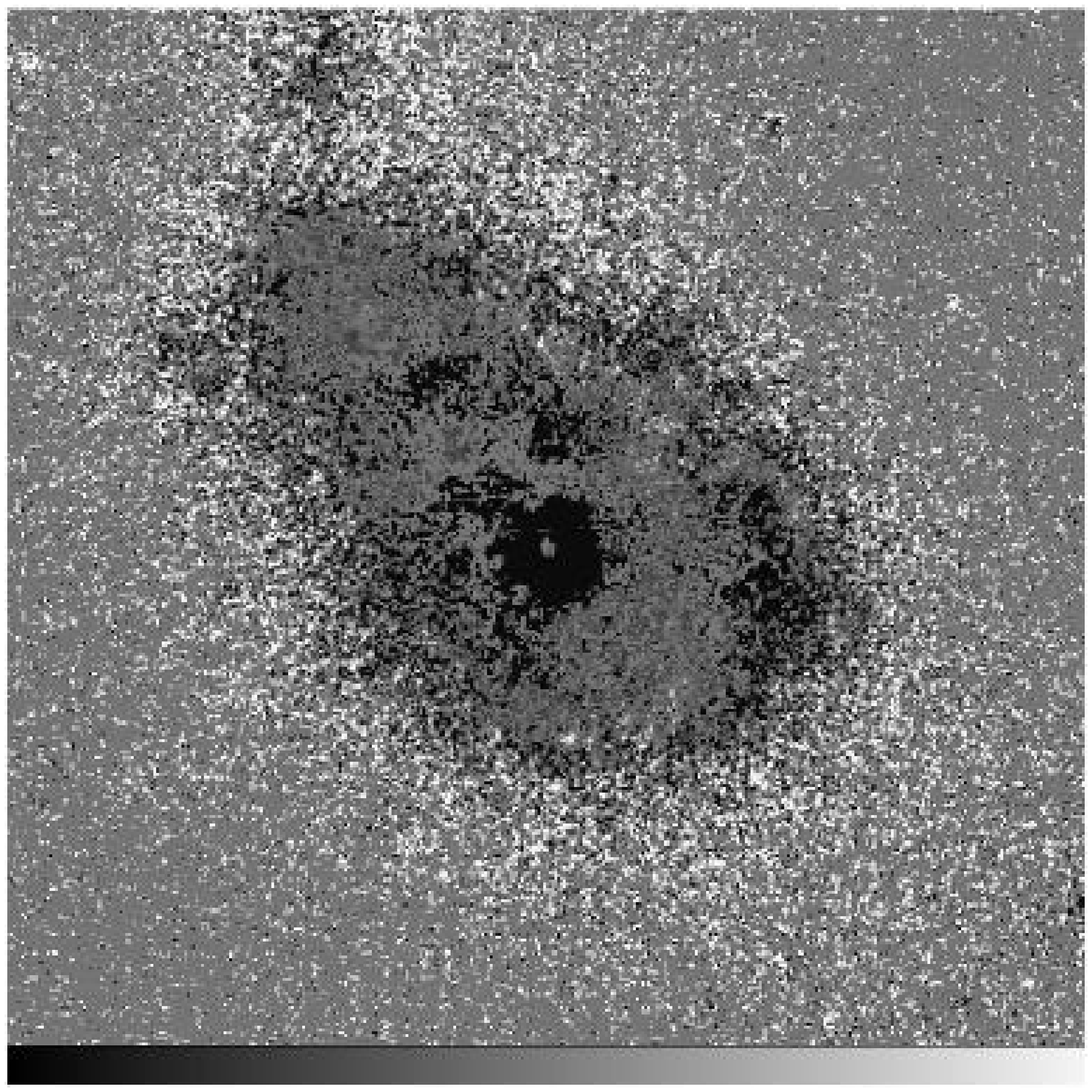}}}
 \resizebox{0.35\hsize}{!}{\rotatebox{-90}
                             {\includegraphics{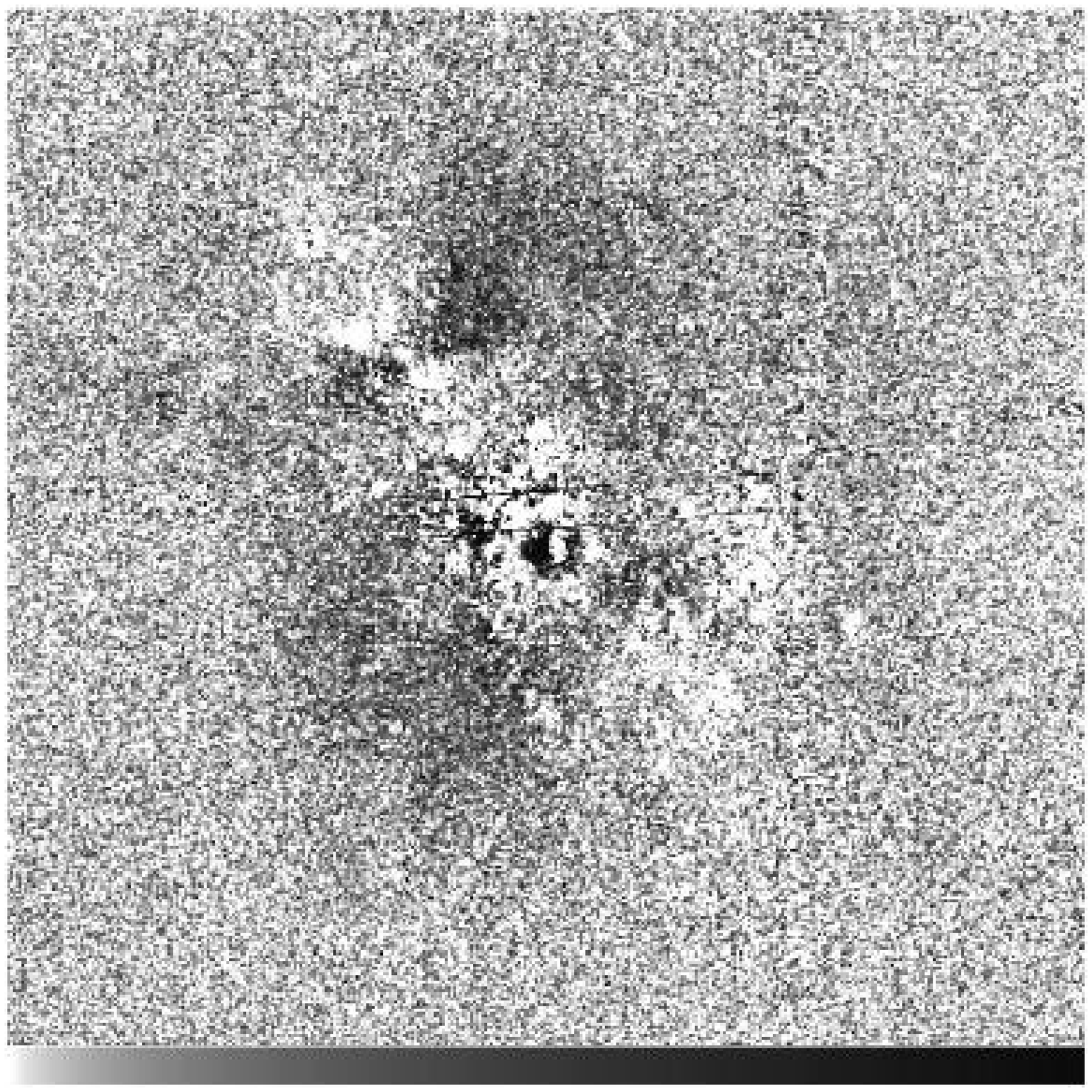}}}
   \caption{{\bf Left}: Scale factor map, ramp scale.
                        Cuts levels are set to show the interesting
                        intensity levels: lcuts=7 (black) to hcuts=14
                        (white).
            {\bf Right}: Continuum subtracted \lyal\  image, created using 
                         the scale factor map. 
           }
      \label{FigResult}
\end{figure}

\medskip
A scale factor map and associated continuum subtraction are shown in 
Fig ~\ref{FigResult}. 
The first continuum subtracted \lyal\  image shows excellent
agreement with features in the spectrum. \lyal\  is seen in both emission
and absorption across the starbusrt region. Emission is seen from knot A
and the region surrounding knot C with a P-Cygni profile at knot A.      
Diffuse emission is seen around knot D but not emission from this knot itself. 
Diffuse emission is also visible in large regions outside starburst regions.
These are \lyal\ photons produced elsewhere that, after multiple 
scatterings, manage to leave the cloud thanks to a low dust content. 
Little is seen around knot B which is really strong
in continuum but shows almost nothing in the low-res STIS spectrum.

We are now in the process of refining this procedure and will apply it
to our whole sample using available HST images. In principle the method
is applicable also to ground based data of high-$z$ \lyal\  galaxies.

\section{Implications for high-redshift galaxies}

In many respects, \lyal\ could be a fundamental probe of the young universe.
It suffers from fewer luminosity biases than Lyman-break techniques so that
\lyal\ surveys become a more efficient way to trace the fainter end of the
luminosity function, i.e., it traces the building blocks of present-day
galaxies in the hierarchical galaxy formation paradigm (Hu, Cowie, \&
McMahon 1998; Fynbo et al. 2001). Early results paint a complex
picture. The equivalent widths of the sources are much larger than expected
for ordinary stellar populations (Malhotra \& Rhoads 2002). They could be
explained by postulating an initial mass function (IMF) biased towards more
massive stars, as predicted theoretically for a very metal-poor stellar
population (Bromm, Coppi, \& Larson 2001).  The combined effect of low
metallicity and flat IMF, however, can only partly explain the anomalous
equivalent widths. Additional mechanisms must be at work (AGN activity?). 
Could  spatial 
offsets between the escaping
\lyal\ and the stellar light, together with higher extinction of the dust
shrouded stars be important?
Our HST spectroscopy and recent ACS imagery cautions against using the 
\lyal\ equivalent width as
a star-formation indicator in the absence of spatial information. 

The source numbers themselves are only about 10\% of the numbers expected
from an extrapolation of the Lyman-break luminosity function. Malhotra \&
Rhoads  speculate if the youngest galaxies are preferentially
selected, whereas older populations are excluded, the results are skewed
towards large \lyal\ equivalent widths.  Could dust formation after
$10^7$~yr destroy the \lyal\ photons? The results in the low-redshift
universe suggests otherwise. We find little support for dust playing a {\em
major} role in destroying \lyal\ photons. Rather, the ACS images favor the
complex morphology as key to understanding the \lyal\ escape mechanism.

Haiman \& Spaans (1999) proposed \lyal\ galaxies as a direct and robust test
of the reionization epoch (see Schaerer, this conference). Prior to 
reionization, these galaxies are hidden
by scattering of the neutral intergalactic medium (IGM). Therefore, a
pronounced decrease in the number counts of {\em galaxies} should occur at
the reionization redshift, independent of Gunn-Peterson trough observations
using {\em quasars}. \lyal\ in galaxies would have the additional appeal of
being sensitive at much higher IGM optical depths since its red {\em wing}
coincides with the red damping wing of intergalactic \lyal\ only. While this
idea is attractive in principle, our ACS imagery calls for caution. In
practice, \lyal\ is a complex superposition of emission and absorption in
the star-forming galaxy itself.  The resulting \lyal\ profile will be
strongly affected by absorption from the continuum. This becomes even more
of a concern when imaging data are interpreted.  When integrating over the
filter bandpass, the emission part of the profile is partly compensated by
the absorption.  As a result, we measure significantly lower \lyal\ fluxes
than with spectroscopic methods, and the escape fraction of \lyal\ photons
is significantly underestimated.

The \lyal\ line is a premier star-formation tracer, in particular at high $z$
where traditional methods, such as H$\alpha$ or the far-IR emission, 
become impractical. Radiative transfer effects in the surrounding 
interstellar gas make its interpretation in terms of star-formation 
rates less straightforward than often assumed. Clearly,
a better understanding of the complex \lyal\ escape mechanisms, both
empirically and theoretically, is required before we can attempt
to interpret large-scale \lyal\ surveys. Future high-resolution imagery of 
local starburst galaxies may provide the necessary calibrations.

\end{document}